\documentclass{article}
\usepackage{emulateapj}
\input epsf.tex
\begin{document}
\title{WHITE DWARFS IN NGC~6791: AVOIDING THE HELIUM FLASH}
\author{Brad M. S. Hansen\altaffilmark{1}
}
\affil{Department of Physics \& Astronomy, University of California Los Angeles, Los Angeles, CA 90095, hansen@astro.ucla.edu}
\altaffiltext{1}{Alfred P. Sloan Foundation Fellow}

\slugcomment{\it 
}

\lefthead{Hansen }
\righthead{NGC6791 White Dwarfs}

\begin{abstract}
We propose that the anomalously bright white dwarf luminosity function observed in NGC~6791 (Bedin et al 2005) is the consequence of the formation of $0.5 M_{\odot}$ white dwarfs with Helium cores 
instead of Carbon cores. This may happen if mass loss during the ascent of the Red Giant Branch is strong enough to
prevent a star from reaching the Helium flash. 
 Such a model can explain the slower white dwarf cooling (relative to standard models)
and fits naturally with scenarios advanced to explain Extreme Horizontal Branch stars,
a population of which are also found in this cluster.

\end{abstract}

\keywords{open clusters and associations: individual (NGC~6791) -- stars: white dwarfs, luminosity function, mass loss, horizontal branch, evolution}

\section{Introduction}

The open cluster NGC~6791 is an interesting object for several reasons. It is amongst the richest
of the Galactic disk clusters, possibly the oldest, and possesses a significantly supersolar 
metallicity. These
attributes have made it a target of several detailed studies. Nevertheless, there is still 
some uncertainty as to the true age. Stetson et al (2004) give an estimate $\sim 12$~Gyr, while
studies such as that of Carney, Lee \& Dodson (2005) or King et al (2005) yield estimates
closer to $\sim 8$~Gyr. The age uncertainties are covariant with uncertainties
in the cluster metallicity ($\rm \left[ Fe/H \right] \sim 0.25$--$0.5$) and distance.

The above estimates are all arrived at by measurements of the main sequence turnoff and
giant sequences. The detection of a significant white dwarf population in NGC~6791 (Bedin et al 2005)
is thus of considerable interest, since it offers the potential to measure cluster parameters
by entirely independent means. However,
 fitting standard white dwarf models to the observed luminosity function yields a cluster age
of $\sim 3$~Gyr.
 In this paper we examine some non-standard white dwarf models
and propose an explanation for this puzzling discrepancy.

In \S~\ref{Models} we review the results of Bedin et al and summarise the various possible 
explanations
they were able to rule out. Thereafter we describe two additional possibilities
not considered in that paper
(residual nuclear burning and the production of massive white dwarfs with Helium cores).
In \S~\ref{LF} we compare these models to the data and demonstrate that the latter may indeed
provide an explanation for the NGC~6791 luminosity function. In \S~\ref{Discuss} we examine
some of the consequences of this model and possible predictions.

\section{The White Dwarf Luminosity Function}
\label{Models}

The luminosity function derived by Bedin et al has the characteristic
shape expected for the white dwarf population drawn from a burst of star formation
-- strongly peaked near a limiting value followed by a sharp drop at lower luminosities.
The problem is that the location of the peak is far brighter than was anticipated.
The expected absolute magnitude for a $0.5 M_{\odot}$ white dwarf 
of age 8~Gyr is $M_{606} \sim 16$ which, combined with a distance modulus of
$\mu_{606}=13.44$, yields an expected peak location of $F606W=29.5$. Yet the
luminosity function of Bedin et al peaks at approximately $F606W\sim 27.5$. 
 They
present compelling evidence that the peak lies well above their completeness limit,
so that one is forced to consider ways to generate a peak in the luminosity function
at much brighter magnitudes than expected.

Bedin et al reviewed several possible explanations for their 
luminosity function. Simply decreasing the distance or extinction is not viable, as
it would drive the age derived from the  main sequence turnoff to unacceptably large
values. Small changes in the white dwarf models, such as changing the Hydrogen layer
thickness or the ratio of Carbon and Oxygen in the core, do not produce 
 a big enough effect. White dwarfs produced by truncated stellar evolution
 in binaries (Kippenhahn, Kohl \& Weigert 1967) do cool more slowly, 
but generally have a range of masses $\sim 0.3$--$0.4 M_{\odot}$ and so would be redder
 than the observed white dwarf cooling track.
Indeed, any explanation involving binary stars needs to account for the narrow mass range
and the lack of bright (relative to a white dwarf) companions at the present time.

There is one potential explanation in the literature, that actually predates the Bedin et al observation.
Bildsten \& Hall (2001) and
Deloye \& Bildsten (2002) discuss the retardation of white dwarf cooling that is made possible
by the sedimentation of $^{22}$Ne during the cooling process. Although the contribution is
small for most white dwarfs, it gets larger if there is more $^{22}$Ne -- which is expected
from more metal-rich systems. Thus, Deloye \& Bildsten in fact predicted that the effects
of sedimentation would be largest in such a system as NGC~6791. At first glance, the NGC~6791
white dwarfs might appear to be a stunning confirmation of the prediction of `sedimentars', but
the size of the predicted effect is somewhat smaller than that needed to explain the Bedin
et al result. Figure~12 and 13 of Deloye \& Bildsten can be converted into a prediction that
the bulk of the white dwarfs should be found between $F606W=28$--$29$, depending on the value
of the assumed diffusion coefficient (the brightest value being found for a rate ten times
faster than the nominal but uncertain value). Nevertheless, the agreement may be improved
by further calculations in progress (L.Bildsten,
personal communication).

At present, however, the observations still require an explanation  
and so we wish to re-examine two
issues touched on by Bedin et al, but not explored to the fullest.

\subsection{Thick Hydrogen Layers and Nuclear Burning}

Bedin et al did examine the effects of changing the Hydrogen layer mass on the
cooling, but as far as can be told from the paper, considered only the effect
it has on heat transport from the core to the surface. For a large enough Hydrogen
layer, the pressure and temperature at the base are high enough to maintain some
level of residual nuclear burning by the pp process, i.e. an additional
heat source. This mechanism has been invoked
in previous instances of anomalously young white dwarfs with 
low mass (e.g. Alberts et al 1996; Driebe et al 1998).
 Could residual nuclear burning be the explanation for the bright white dwarfs in 
NGC~6791? Unfortunately, this seems unlikely. The
effects of nuclear burning are only significant for masses $M<0.25 M_{\odot}$ (Driebe et al 1999;
Althaus, Serenelli \& Benvenuto 2001; 
Hansen, Kalogera \& Rasio 2003).

 To confirm this, we calculate several new white
dwarf cooling sequences using the code described in Hansen (1999).
The Helium layer is taken to be $q_{He} = 10^{-2}$. All models have total
mass $0.5 M_{\odot}$.
 We use the same
C/O profile as in the previous models, but consider a variety of Hydrogen layer
thicknesses, ranging from $q_{H}=10^{-4}$ to $q_{H} = 5 \times 10^{-3}$. The 
lower end of the range is the canonical value expected for most white dwarfs,
but larger values are possible given the uncertain nature of the mass loss
history of evolved stars. 

In the initial stages (central temperatures $>10^8$ K) of the approach to the cooling sequence, the evolution
is driven by neutrino cooling in the core and the cooling is similar for all models.
However, as the star shrinks towards a more compact, degenerate configuration, the 
pressure and temperature at the base of the Hydrogen layer increase and nuclear burning
is possible for some range of Hydrogen layer masses. For even the most massive
layers considered, the models
 remain brighter than the fiducial ($q_H = 10^{-4}$) models only for $\sim 2$~Gyr.
This is because, if
one increases the mass of the Hydrogen layer, the pressure and density at the
base of the envelope are higher and so the rate of burning is higher. In the
end the effect of a  larger reservoir is balanced by a higher rate of consumption,
limiting the delay one can achieve. After 2~Gyr, no model has a Hydrogen
layer larger than $q_H \sim 5 \times 10^{-4}$.
Thus, after $\sim 8$~Gyr, the white dwarf will have approximately the
same luminosity regardless of the initial Hydrogen mass. This is demonstrated in
Figure~\ref{LT}. The white dwarf cooling time (neglecting the main sequence lifetime
of the progenitor for now) for white dwarfs at $F606W \sim 27.3$ is 1.6~Gyr for
a standard C/O model. Even the thickest Hydrogen envelope considered ($q_H \sim 5 \times 10^{-3}$)
only lengthens the cooling time to this luminosity by 0.6~Gyr. 

\subsection{Helium Cores}
\label{HeCore}

The notion that the NGC~6791 white dwarfs have cores composed of Helium, rather
than Carbon or Oxygen, was touched on briefly by Bedin et al. However, they ruled
this out as Helium core white dwarfs are believed to result from binary evolution
and the majority 
 have masses that range $\sim 0.3$--$0.4 M_{\odot}$ and would thus be too
red to fit the observations.

We believe this to be an overly conservative restriction. In this section we examine
models for white dwarfs of mass $0.45$--$0.55 M_{\odot}$ (models with this mass
are consistent with the colour and magnitude of the upper cooling sequence)  which nevertheless possess
cores composed purely of Helium. We will leave issues of provenance to \S~\ref{LF}.

For white dwarfs hot enough to be far from the strongly coupled regime
in the core, the heat content (and so the cooling time) of the star is inversely proportional to the mass number of
the core constituent. Thus, if we consider two $0.5 M_{\odot}$ white dwarfs, one composed of Carbon
and the other of Helium, both with an age 8~Gyr, then the latter will be considerably brighter.
In fact, it will have approximately the same luminosity as the Carbon core model has at an 
age $\sim 8/3 \sim 2.7$~Gyr. This is approximately the age 
 inferred by Bedin et al. 
Figure~\ref{LT} shows that the Helium core model fares considerably better than the C/O models,
 yielding a cooling
time of $\sim 4$~Gyr to $M_V\sim 27.3$. While this is still somewhat below the 8~Gyr age of the
cluster, we must recall that the cluster age is the sum of the time spent on both the cooling
sequence and on the main sequence. Thus, the difference between the cluster age and the Helium-core
cooling time indicates the mass of the main sequence progenitors of this anomalous population.

We use the analytic stellar evolution formulae encoded in the {\tt SSE} package (Hurley, Pols \& Tout 2000) to calculate the main sequence lifetimes
of stars with $Z=0.035$. For the low end of the cluster age range ($\sim 8$~Gyr), the progenitor
mass for Helium core white dwarfs at the observed cutoff
 is $\sim 1.45 M_{\odot}$, while for the high end of the age range ($\sim 12$~Gyr), the
progenitor mass  is $\sim 1.17 M_{\odot}$. 
Exploring the range of applicable white dwarf masses
($0.45 M_{\odot}$ to $0.55 M_{\odot}$) does not change the conclusions significantly.

Next we will try to model the luminosity function. Before doing this we need to consider in
more detail the
possible origins of such a population.

\section{Avoiding the Helium flash in NGC~6791}
\label{LF}

There are good reasons to believe most white dwarfs have C/O cores.
Upon reaching the tip
of the Red Giant Branch (RGB), a star 
ignites core Helium burning (under degenerate conditions for lower mass progenitors -- which
leads to a thermonuclear runaway, a.k.a. the Helium flash), and the star moves onto the Horizontal Branch.
It undergoes extended core Helium burning followed by  shell burning 
 on the Asymptotic Giant Branch before
becoming a white dwarf. In order to avoid converting the Helium core to Carbon and Oxygen the
star has to find some way to circumvent these later stages of stellar evolution.

Stars in binaries sometimes manage this feat by losing their envelope on the ascent
of the RGB due to Roche lobe overflow i.e., mass transfer to a companion.
However, this leads to a range of masses, most of which are considerably lower than
the $\sim 0.5 M_{\odot}$ needed to fit the NGC~6791 cooling sequence. Bedin et al
have already considered and rejected this possibility. In order for our model to
work we have to hypothesize that many single stars in NGC~6791 have managed to
lose their envelopes before reaching the Helium flash, so that they move directly
from the RGB to the white dwarf cooling sequence.

In fact, this
 hypothesis dovetails quite naturally with several theories for the origins and
nature of Extreme Horizontal Branch (EHB) stars. This term is used to describe a
class of stars, found both in the field and in clusters, which appear to be
related to traditional Horizontal Branch stars but which are hotter/bluer.
 It has been suggested (Faulkner 1972; Sweigart, Mengel \& Demarque 1974) that 
such stars are core Helium burning stars with particularly thin Hydrogen envelopes,
possibly as a result of extreme mass loss on the RGB. Subsequent work has developed
this picture even further.
Castellani \& Castellani (1993) and
Castellani,
Luridiana \& Romaniello (1994) report the formation of what they call ``Red Giant Stragglers''
in models for globular cluster evolution. In some cases, the stars make it to the white dwarf
cooling sequence before igniting Helium. 
D'Cruz et al (1996) considered models with a range of mass loss rates on the RGB and found
that, in addition to the formation of EHB stars, they formed
so-called ``flash-manque'' stars, which lose so much mass that they never ignite Helium and
simply go directly 
 to Helium core white dwarfs. In fact, given the large amount of mass loss required to
form EHB stars, especially the very bluest `blue hook stars' (Brown et al 2001;
Cassisi et al 2003; Moehler et al 2004),
it would require extreme fine tuning to produce EHB stars without also producing some
Helium core white dwarfs.
This is important because NGC~6791 possesses a significant population
of such EHB stars (Kaluzny \& Udalski 1992; Liebert, Saffer \& Green 1994). Thus, if one
takes the above models at face value, a substantial population of Helium core white
dwarfs is expected wherever EHB stars are found and thus should be found in NGC~6791.

We now try to model the observed luminosity function using this framework. 
We note that not all the white dwarfs in NGC~6791 can have Helium cores.
 The models discussed above that
successfully avoid the Helium flash do so with progenitors that begin their lives
with masses slightly larger than $1 M_{\odot}$. Thus, in our models we shall impose a critical
mass $m_{crit}$, above which stars always produce standard C/O core white dwarfs.
 Furthermore,
 NGC~6791 does possess normal Helium-burning stars (the EHB stars make up 
$\sim 15\%$ of the Helium-burning stars according to Liebert et al 1994) and even the EHB stars are Helium-burning, so
that clearly some C/O white dwarfs are being produced. 
It seems likely that stellar evolution in this cluster explores all three post-RGB
avenues discussed above. One may estimate the branching ratios as follows.

King et al (2005) 
 note that two of
the EHB candidates from Kaluzny \& Udalski (1992) lie within the HST field. 
Models suggest that this evolutionary stage lasts for
 $\sim 10^8$ years. The offspring of EHB stars are C/O white dwarfs and
so, using the cooling time of C/O models (to $F606W=28$), we estimate that 
 $\sim 2 \times 10^9/10^8 \times 2 = 40$ of the
white dwarfs in this field brighter than $F606W=28$ should have come through this channel.
In addition, C/O white dwarfs are also the end product of normal Helium-burning stars
in the cluster. Although we do not have a strict count of HB stars in this field,
 Liebert et al (1994) estimate that
$\sim 15$\% of all NGC~6791 Helium-burning stars are EHB stars. Thus, we
estimate that the total number of
C/O white dwarfs produced in this field with $F606W<28$ is $\sim 40/0.15\sim 270$.
The total number of white dwarfs observed by Bedin et al in the same field is $\sim 600$.
The difference between these two numbers is the number of Helium core dwarfs that
went directly from RGB to cooling sequence.
Thus, the ratio of C/O core white dwarfs to Helium core dwarfs is estimated to be $\sim 270:330 \sim 5:6$.
If we break this up further into EHB/normal~HB/direct He core we estimate ratios $\sim 1:6:8$.
The principal uncertainty in this procedure is the degree to which internal cluster dynamical evolution
violates the implicit closed box assumption. Although EHB stars and white dwarfs
should have very similar masses, the normal HB stars may have slightly higher masses and thus
may be slightly underepresented (since the field is away from the cluster core)
 in the census just described. In light of this uncertainty,
we adopt a ratio of C/O to Helium cores of 1:1 below.

Figure~\ref{LF1} shows the luminosity function of Bedin et al (now binned in 0.5 magnitude
bins) compared to sample 8~Gyr model luminosity functions. The best fit value of $m_{crit}=1.6 M_{\odot}$
is slightly larger than the $1.45 M_{\odot}$ quoted in \S~\ref{HeCore} because the peak
of the realised Monte Carlo number distribution lies slightly below the cutoff. The
solid line shows a histogram for a model in which all stars with $m<m_{crit}$ form Helium
core white dwarfs. The dotted line shows the C/O core luminosity function for the same
parameters. The dashed line corresponds to a model in which 50\% of the stars with
$m<m_{crit}$ form Helium cores and the rest form C/O cores. We have included the effects
of incompleteness, using the results from Bedin et al, kindly provided by Ivan King. The truncation
of the C/O core luminosity function shows where this becomes important. The truncation of
the Helium core luminosity function occurs at significantly brighter magnitudes. 
The uncertain age of the cluster means that there is some flexibility in the
choice of parameters. Figure~\ref{LF2} shows a fit using an age of 12~Gyr and
$m_{crit}=1.25 M_{\odot}$. The solid, dotted and dashed lines once again indicate
luminosity functions for Helium core, C/O core and a 1:1 combination, as in
Figure~\ref{LF1}.

There are several points to note about these fits. We have not performed a detailed parameter
scan in fitting the data because we do not possess a proper photometric uncertainty map for
this data (which indicates the probabilistic relation between intrinsic and observed magnitude as
a function of model magnitude -- see Hansen et al (2004) for a more detailed discussion of
this question in the context of the globular cluster M4).
The structure in the CMD in Figure~1 of Bedin et al suggests this is likely to be an
important consideration in performing a true fit. We also need to assume something about
the distribution of progenitor masses. We have assumed top heavy mass functions, with 
a slope $x<-1$ (where Salpeter is $x=1.35$).
This is necessary to obtain luminosity functions as peaked as those
seen here, and is also broadly consistent with the main sequence mass function observed
by King et al (2005) for this cluster (the number of main sequence stars per magnitude
bin increases with decreasing magnitude/increasing mass near the turnoff). For an open
cluster as old as NGC~6791, there has likely been significant dynamical evolution and so
one should be cautious when interpreting this as a true mass function.

\section{Discussion}
\label{Discuss}

In this paper we propose a model to explain the origin of the peak in the NGC~6791 
luminosity function at luminosities somewhat brighter than those expected on the
basis of traditional white dwarf cooling models. Our proposal is that these are
the result of the formation of Helium core white dwarfs resulting from strong
mass loss on the RGB, thereby avoiding an episode of core Helium
burning on the Horizontal Branch. In support of this model we point to the substantial
population of EHB stars in NGC~6791, which are proposed to be
stars which lost enough mass to almost make it to the white dwarf cooling track 
with their Helium cores intact but finally ignited Helium burning during the
final contraction stage. In many models, EHB stars and Helium core white dwarfs
are expected to form together as they are consequences of the same phenomenon --
strong mass loss.

The scenario discussed here does make a definite prediction -- that the white dwarf
luminosity function for NGC~6791 should be bimodal, with a second peak at fainter
magnitudes resulting from traditional C/O white dwarfs which did, indeed, pass
through the Horizontal Branch phase. Figure~\ref{LF3} shows the expected luminosity
function for our simple 8~Gyr model with $m_{crit}=1.6 M_{\odot}$.
The solid, dotted and dashed histograms show the same luminosity functions as in
Figure~\ref{LF1} but now we have not modelled the effects of incompleteness.
We see that the second peak is expected to lie at magnitudes $F606 \sim 29$--$30$.
The exact value will vary depending on the mass of the white dwarfs at the
faint end. The value shown here is probably a little optimistic since we used only
$0.5 M_{\odot}$ white dwarfs. 


The narrowness of the white dwarf cooling sequence suggests that, however these  Helium-core white dwarfs
form, it must be by a process that strongly favors core masses $>0.4 M_{\odot}$. If heavy mass loss
on the RGB is indeed the cause, this mass limit suggests that the progenitors must get within
$\sim$~1~magnitude of the tip of the RGB before their evolution is truncated (based on metal-rich
models using the formulae of Hurley et al 2000). There is some observational support for this
notion. Origlia et al (2002) used ISOCAM to search for infrared excesses suggestive of mass
loss in the cores of five globular clusters. They found evidence for significant mass loss only
close to the RGB tip. Similarly, Ita et al (2002) found many variable stars near the tip of the RGB
in the Large Magellanic Cloud, possibly an indication of pulsation driven mass loss. While these
results are encouraging, they only indicate that most of the mass loss on the RGB is indeed likely
to occur near the tip. 
 There is clearly still much work required to verify that the amount of mass loss on
the NGC~6791 RGB is sufficient to justify our hypothesis.

Finally, the referee has raised the question whether the existence of this anomalous white
dwarf population poses a problem for the determination of population ages from the white
dwarf cooling sequence (Fontaine, Brassard \& Bergeron 2001, Hansen 2004 and references therein).
 Within the model proposed here, this is not the case, because the difference between
the Helium core and Carbon core white dwarfs is discrete in nature and any anomaly
is readily apparent. Just as someone cannot be ``a little bit pregnant'', a star must belong
to one population or the other.  
If the model of Deloye \& Bildsten is correct, however, then the anomaly is ``tunable'' 
via the abundance of $^{22}$Ne and thereby does indeed introduce an additional uncertainty
to take into account. Thus, it is of considerable interest to distinguish between these
two possibilities with a deeper observation of this cluster.

\acknowledgements
The work described here is supported by NASA grant ATP03-000-0084 and the Alfred P. Sloan
Foundation. The author acknowledges helpful discussions with Jason Kalirai, Mike Rich,
Harvey Richer, Ivan King, Francesco Ferraro and Lars Bildsten. He thanks the referee for a 
thoughtful and helpful referee report.

\newpage

\figcaption[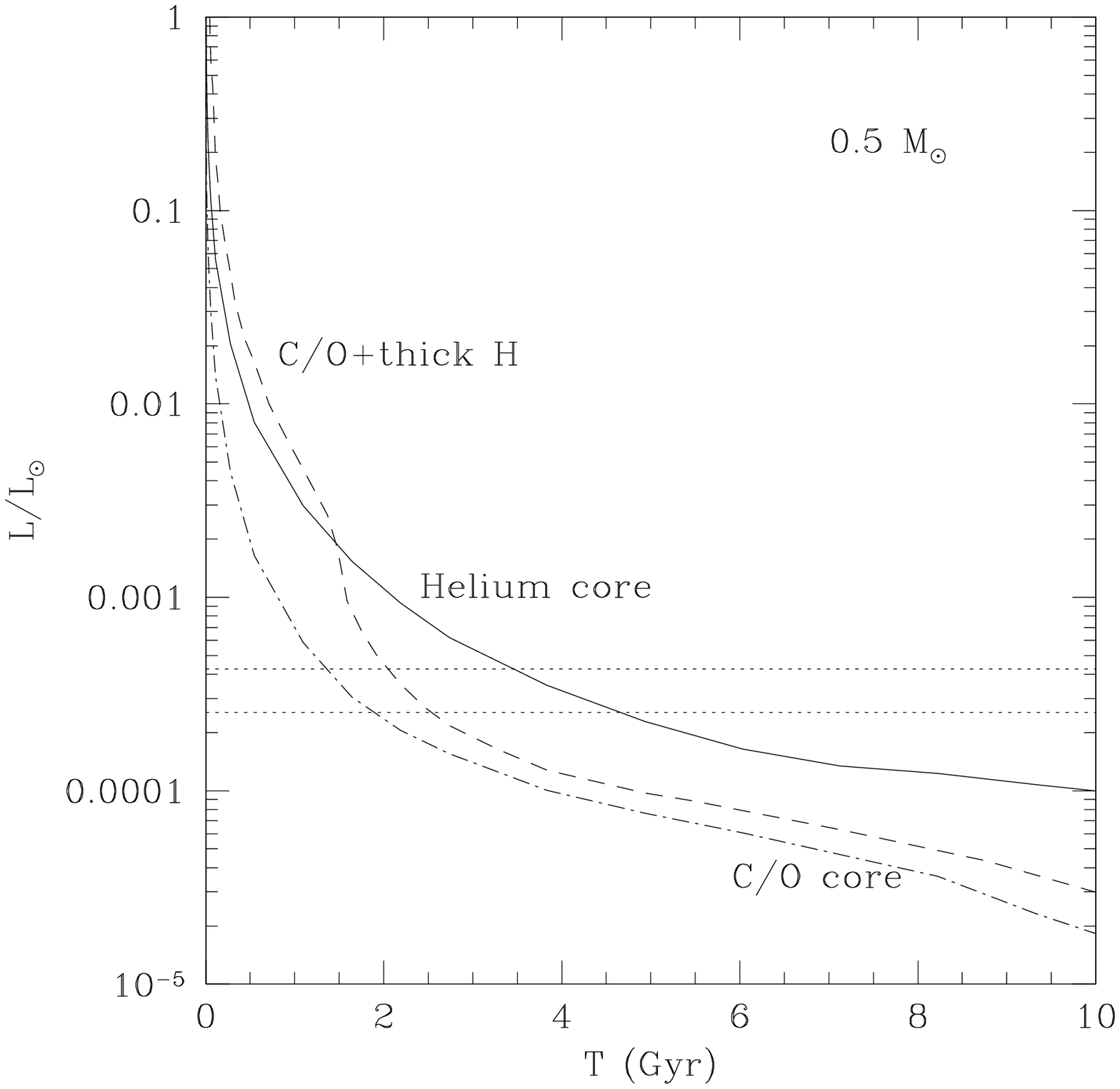]{ The horizontal dotted lines bracket the luminosity range 
corresponding to the Bedin et al peak. We have used the distance modulus $M_{606}=13.44$
and colour--$\rm T_{eff}$ transformations from Bergeron, Wesemael \& Beauchamp (1995). The dot-dashed curve corresponds
to a standard 0.5$M_{\odot}$ white dwarf with C/O core and normal Helium and Hydrogen
layers ($q_{He}=10^{-2}$, $q_H=10^{-4}$). The dashed curve is for the same model but now
with a much thicker Hydrogen layer $q_H=5 \times 10^{-3}$. This results in a period of
nuclear burning during which the star is considerably brighter. However, the Hydrogen
fuel is exhausted after $\sim$2~Gyr. The solid curve shows a $0.5 M_{\odot}$ model with
pure Helium core and a Hydrogen layer envelope of $q_H=10^{-4}$. The cooling is $\sim 2.5$
times slower because of the higher total heat capacity of the Helium core.
 \label{LT}}

\figcaption[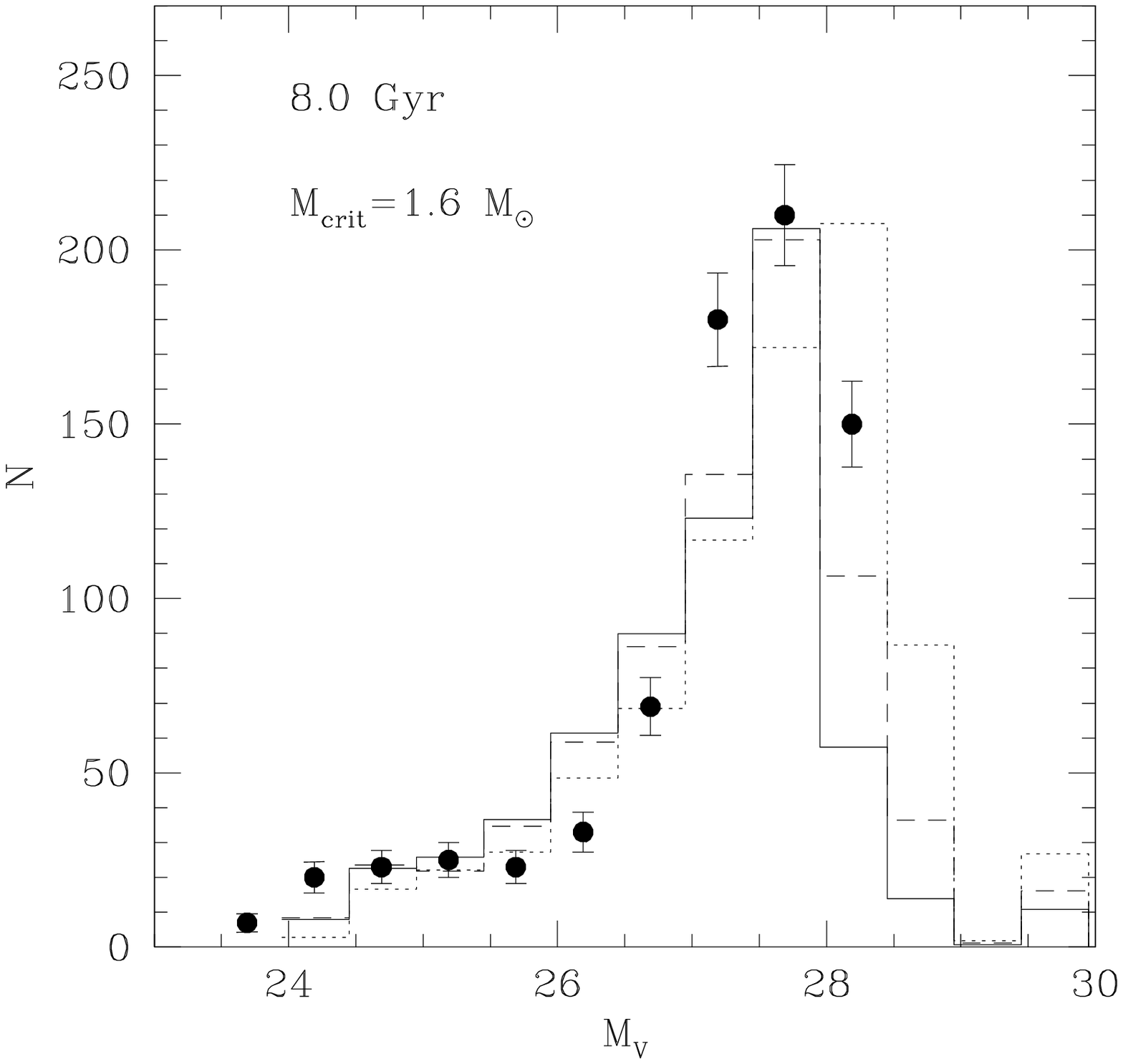]{
The points show the observed
luminosity function from Bedin et al. (2005). This is compared
to three theoretical luminosity functions. All are calculated for
a cluster age of 8~Gyr. The solid luminosity function represents
a population for which all stars with $m<m_{crit}=1.6 M_{\odot}$ for
Helium core white dwarfs. The dotted luminosity function is the 
standard model in which all stars produce C/O core white dwarfs.
The dashed luminosity function is one in which only 50\% of
the stars with $m<m_{crit}$ make Helium core white dwarfs. 
The theoretical models include a correction for the incompleteness
of the Bedin et al observations. For simplicity, in this model
all white dwarfs are assumed to have mass $0.5 M_{\odot}$.
 \label{LF1}}

\figcaption[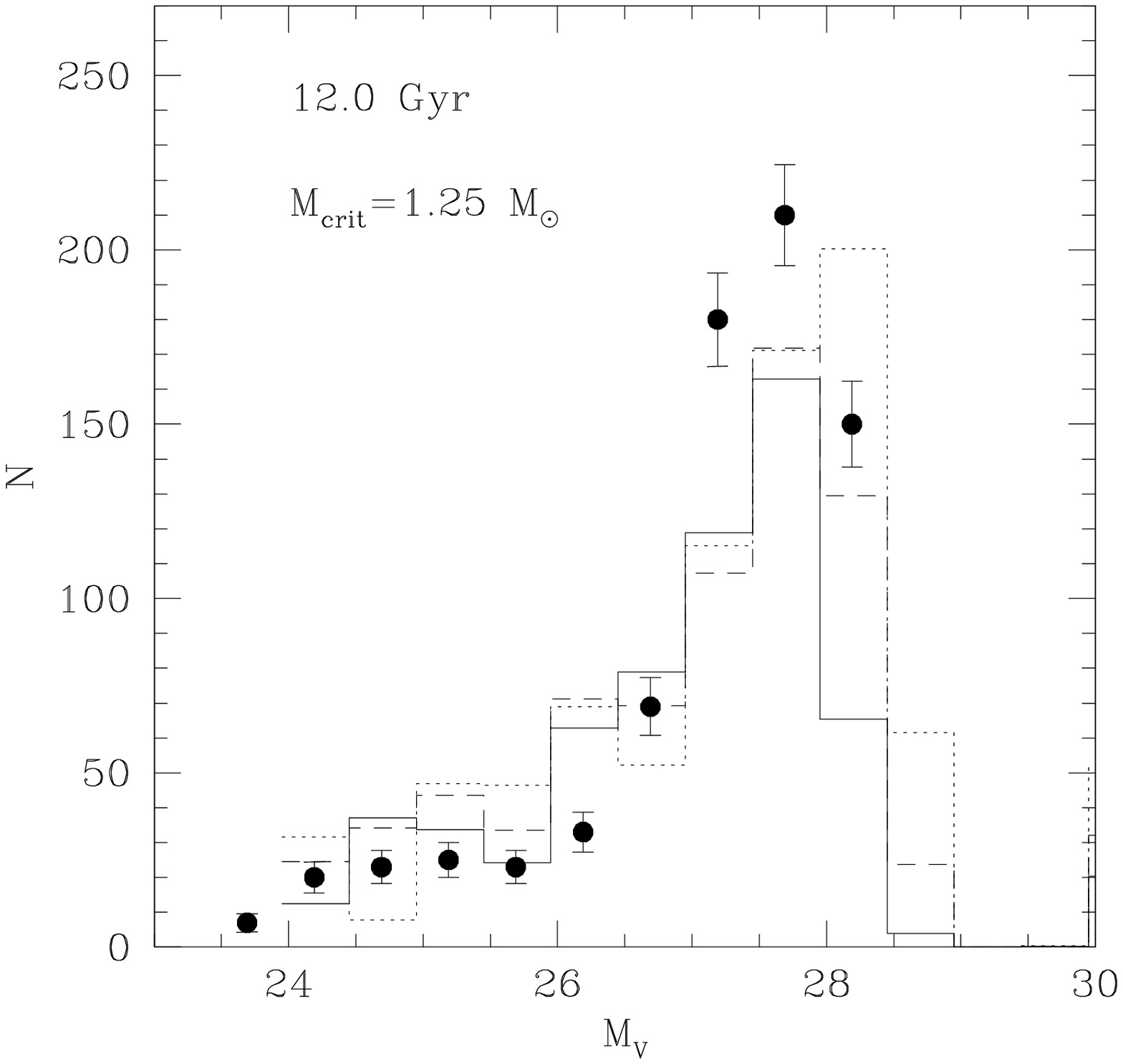]{ The format for this model is the same 
as for Figure~\ref{LF1}, except that the cluster age is now
assumed to be 12~Gyr and the value of $m_{crit}=1.25 M_{\odot}$.
 \label{LF2}}

\figcaption[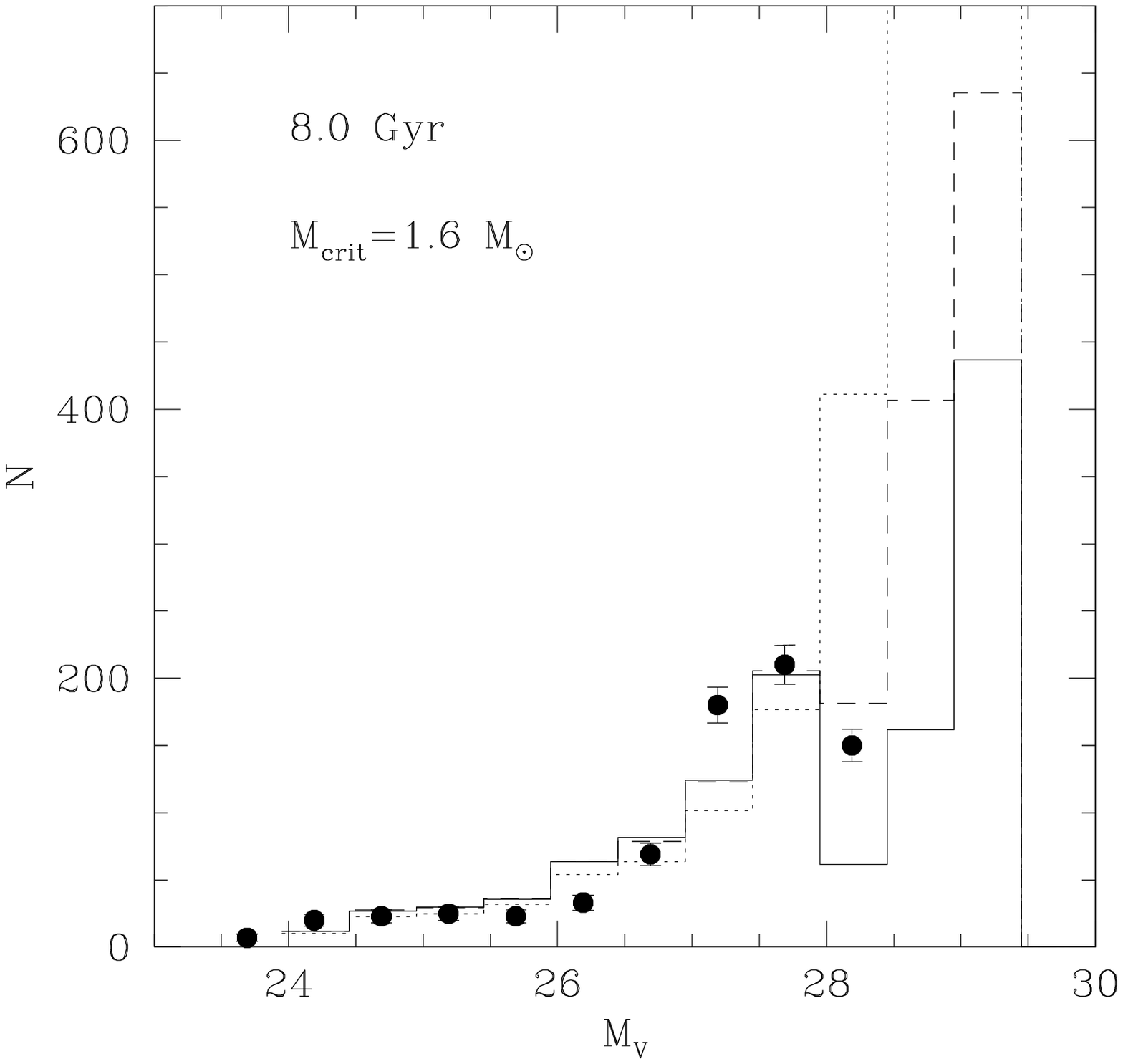]{The format for this figure is once again
that of Figure~\ref{LF1}, including the age and $m_{crit}$,
except that now we make no correction for incompleteness. This
shows the large number of white dwarfs we expect to find at
fainter magnitudes if this model is correct.
 \label{LF3}}

\clearpage
\plotone{f1.eps}
\clearpage
\plotone{fig2.ps}
\clearpage
\plotone{fig3.ps}
\clearpage
\plotone{fig4.ps}

\begin{references}
\reference{Alb} Alberts, F., Savonije, G. J., van den Heuvel, E. P. J. \& Pols, O. R., 1996,
Nature, 380, 676
\reference{ASB} Althaus, L. G., Serenelli, A. M. \& Benvenuto, O. G., 2001, ApJ, 554, 1110
\reference{Bedin} Bedin, L. R., Salaris, M.,  Piotto, G., King, I. R., Anderson, J.,
Cassissi, S. \& Momany, Y., 2005, ApJ, 624, L45
\reference{BWB} Bergeron, P., Wesemael, F. \& Beauchamp, A., 1995, PASP, 107, 1047
\reference{BH} Bildsten, L. \& Hall, D. M., 2001, ApJ, 549, L219
\reference{BWL} Brown, T. M., Sweigart, A. V., Lanz, T., Landsman, W. B. \& Hubeny, I., 2001, ApJ, 562, 368
\reference{CLD} Carney, B. W., Lee, J.-W. \& Dodson, B., 2005, AJ, 129, 656
\reference{CSSW} Cassisi. S., Schlattl, H., Salaris, M. \& Weiss, A., 2003, ApJ, 582, L43
\reference{CC} Castellani, M. \& Castellani, V., 1993, ApJ, 407, 649
\reference{CLR} Castellani, V., Luridiana, V., \& Romaniello, M., 1994, ApJ, 428, 633
\reference{DCruz} D'Cruz, N.L., Dorman, B., Rood, R. T. \& O'Connell, R. W., 1996, ApJ, 466, 359
\reference{DB} Deloye, C. J. \& Bildsten, L., 2002, ApJ, 580, 1077
\reference{Dr1} Driebe, T., Sch\"{o}nberner, D., Bl\"{o}cker, T.  \& Herwig, F., 1998, A\&A, 339, 123
\reference{Dr2} Driebe, T., Bl\"{o}cker, T., Sch\"{o}nberner, D., \& Herwig, F., 1999, A\&A, 350, 89
\reference{F72} Faulkner, J., 1972, ApJ, 173, 401
\reference{FBB} Fontaine, G., Brassard, P. \& Bergeron, P., 2001, PASP,  113, 409
\reference{Hanm1} Hansen, B., 2004, Phys. Rep., 399, 1
\reference{Han0} Hansen, B., 1999, ApJ, 520, 680
\reference{Han1} Hansen, B. et al., 2004, ApJS, 155, 551
\reference{HKR} Hansen, B., Kalogera, V. \& Rasio, F. A., 2003, ApJ, 586, 1364
\reference{HPT} Hurley, J. R., Pols, O. R. \& Tout, C. A., 2000, MNRAS, 315, 543
\reference{Ita} Ita, Y. et al., 2002, MNRAS, 337, L31
\reference{KU} Kaluzny, J. \& Udalsky, A., 1992, Acta Astron, 42, 29
\reference{King} King, I. R., Bedin, L. R., Piotto, G., Cassissi, S. \& Anderson, J., 2005,
astro-ph/0504627
\reference{KKW} Kippenhahn, R., Kohl, K. \& Weigert, A., 1967, Z. Astr., 66, 58
\reference{LSG} Liebert, J., Saffer, R. A. \& Green, E. M., 1994, ApJ, 107, 1408
\reference{MSL} Moehler, S., Sweigart, A. V., Landsman, W. B., Hammer, N. J. \&
Dreizler, S., 2004, A\&A, 415, 313
\reference{OFF} Origlia, L., Ferraro, F. R., Fusi Pecci, F. \& Root, R. T., 2002, ApJ, 571, 458
\reference{SBG} Stetson, P. B., Bruntt, H. \& Grundahl, F.,2003, 115, 413
\reference{SMD} Sweigart, A. V., Mengel, J. G. \& Demarque, P., 1974. A\&A, 30, 13

\end{references}
\end{document}